\documentclass[letterpaper]{article} 
\usepackage[draft]{aaai25}  
\usepackage{times}  
\usepackage{helvet}  
\usepackage{courier}  
\usepackage[hyphens]{url}  
\usepackage{graphicx} 
\usepackage{natbib}  
\usepackage{caption} 
\usepackage{algorithm}
\usepackage{algorithmic}

\usepackage{newfloat}
\usepackage{listings}
\DeclareCaptionStyle{ruled}{labelfont=normalfont,labelsep=colon,strut=off} 
\floatstyle{ruled}
\newfloat{listing}{tb}{lst}{}
\floatname{listing}{Listing}

\usepackage{tikz}
\tikzset{every picture/.style={line width=0.75pt}} 
\usepackage{subcaption}
\usepackage{latexsym}
\usepackage{amsmath}
\usepackage{amsthm}
\usepackage{amssymb}
\usepackage{mathrsfs}

\usepackage{pdfpages}

\usepackage{style/logic}
\usepackage{style/theorem}

\usepackage{color}

\newtheorem{example}{Example}
\newtheorem{theorem}{Theorem}

\usepackage{cleveref}
\newcommand{\autoref}[1]{\cref{#1}}
\crefname{section}{Section}{Sections}
\crefname{figure}{Figure}{Figures}
\crefname{table}{Table}{Tables}
\crefname{equation}{Equation}{Equations}
\crefname{definition}{Definition}{Definitions}
\crefname{theorem}{Theorem}{Theorems}
\crefname{example}{Example}{Examples}

\usepackage{enumitem}
\usepackage{booktabs}

\title{Traffic Scenario Logic: A Spatial-Temporal Logic for Modeling and Reasoning of Urban Traffic Scenarios}
\author{
    Ruolin Wang\textsuperscript{\rm 1,\rm 2}\equalcontrib,
    Yuejiao Xu\textsuperscript{\rm 1,\rm 2}\equalcontrib,
    Jianmin Ji\textsuperscript{\rm 1,\rm 3}\thanks{Corresponding author.}
}
\affiliations{
    \textsuperscript{\rm 1}University of Science and Technology of China, 230026, China\\
    \textsuperscript{\rm 2}Suzhou Shuzhi Technology Group Co., Ltd., Suzhou 215000, Jiangsu, China\\
    \textsuperscript{\rm 3}Institute of Artificial Intelligence, Hefei Comprehensive National Science Center, 230088, China\\
    \{rl\_wang, yuejiao\}@mail.ustc.edu.cn,
    jianmin@ustc.edu.cn
}

\DeclareMathSizes{7}{7}{7}{6.5}

\begin{document}

\maketitle

\begin{abstract}
Formal representations of traffic scenarios can be used to generate test cases for the safety verification of autonomous driving.
However, most existing methods are limited to highway or highly simplified intersection scenarios due to the intricacy and diversity of traffic scenarios.
In response, we propose Traffic Scenario Logic (TSL), which is a spatial-temporal logic designed for modeling and reasoning of urban pedestrian-free traffic scenarios.
TSL provides a formal representation of the urban road network that can be derived from OpenDRIVE,
i.e., the de facto industry standard of high-definition maps for autonomous driving,
enabling the representation of a broad range of traffic scenarios without discretization approximations.
We implemented the reasoning of TSL using Telingo,
i.e., a solver for temporal programs based on Answer Set Programming,
and tested it on different urban road layouts.
Demonstrations show the effectiveness of TSL in test scenario generation and its potential value in areas like decision-making and control verification of autonomous driving.
The code for TSL reasoning has been open-sourced.
\end{abstract}

\begin{links}
    \link{Code}{https://github.com/USTCAVSA/TSL}
\end{links}

\section{Introduction}

Autonomous driving technology is promising to revolutionize transportation by enhancing safety, efficiency, and accessibility.
The rigorous testing and verifying their safety under various traffic scenarios is central to developing and deploying autonomous vehicles.
Formal representations of these scenarios are crucial for generating comprehensive test cases and ensuring the reliability of autonomous driving systems~\citep{zhang2021test}.

Existing methods for formalizing traffic scenarios often face limitations, particularly in modeling intricate urban environments with diverse traffic patterns.
While some approaches excel in representing highway scenarios or simplified intersections, they struggle to capture the complexity inherent in urban driving environments~\citep{hilscher2016abstract,yao2023hierarchical,li2020ontology}.
Consequently, there is a pressing need for a versatile and robust formal logic framework capable of modeling a broad spectrum of urban traffic scenarios.

In response to this challenge, we introduce \textit{Traffic Scenario Logic (TSL)}, a novel spatial-temporal logic specifically tailored for modeling and reasoning about urban pedestrian-free traffic scenarios.
Unlike previous methods, TSL is designed to provide a comprehensive formal representation of the urban road network without discretization approximations, enabling the modeling of various traffic configurations encountered in complex urban environments.

One of the main benefits of TSL is its compatibility with OpenDRIVE, which is the industry-standard format used for high-definition (HD) maps in autonomous driving~\citep{dupuis2010opendrive}.
By utilizing OpenDRIVE data, TSL seamlessly translates real-world road networks into logical representations, enabling the representation of a wide range of urban traffic scenarios.


TSL also distinguishes itself with a unique spatial representation.
Instead of using a simple discretized position description, TSL captures a vehicle's longitudinal position based on its relationship with other vehicles.
By merging scenarios with the same vehicle relationships, TSL improves efficiency without losing comprehensive coverage of scenarios.

We first give the syntax and semantics of TSL for uni-direction multi-lane highway scenarios in \autoref{sec:HTSL}, then extend it to urban scenarios in \autoref{sec:UTSL}.
Finally, we implement the reasoning of TSL with Telingo~\citep{cabalar2019telingo}, a solver for temporal programs based on Answer Set Programming.
\autoref{sec:application} demonstrates several examples on different urban road layouts, showing the effectiveness of TSL in test scenario generation and potential applications in decision-making and control verification of autonomous driving.

In summary, this paper makes the following contributions:
\begin{itemize}
    \item Introduces Traffic Scenario Logic (TSL) for formal modeling and reasoning of urban pedestrian-free traffic scenarios.
    \item Provides a formal representation of an urban road network that is compatible with the OpenDRIVE HD map format and an efficient and complete spatial representation.
    \item Implements the reasoning mechanism with Telingo solver and provides examples of different road layouts.
\end{itemize}

\section{Related Work}

Constructing a generic spatial-temporal model for the physical world is a challenging and significant field of study.
Prior research has primarily focused on developing different combinations of temporal and spatial logics.
\citet{kontchakov2007spatial} have provided a comprehensive summary of this research. 

\citet{hilscher2011abstract} were one of the first few work on using spatial-temporal logic to model traffic scenarios, proposing \textit{Multi-Lane Spatial Logic (MLSL)} which provides an abstract model of a uni-directional multi-lane motorway.
In MLSL, vehicles' position, speed, and acceleration constitute a traffic snapshot, and the traffic snapshots at different moments constitute a traffic scenario.
Changes between traffic snapshots are described by transition rules,
inspired by $\mathcal{ITL}$~\citep{moszkowski1982temporal}.
\citet{hilscher2013proving} proposed an extended version of MLSL that supports a bi-directional motorway and length measurement for road segments.
\citet{hilscher2016abstract} further proposed \textit{Urban Multi-Lane Spatial Logic (UMLSL)} which contains the modeling of intersections,
simplifying an intersection of two two-way, single-lane highways into four crossing road segments.
Meanwhile, \citet{xu2016spatial} introduced vertical lanes that intersect with the horizontal lanes by extending the position to two dimensions to create an urban road network, formalized in a grid way.
Based on UMLSL, \citet{schwammberger2021extending} proposed \textit{Urban Spatial Logic for Traffic Rules (USL-TR)} to add static objects and non-autonomous road users.
\citet{xu2019scenario} represented an urban road network with road segments and the arcs that link them together,
which was further formalized into a hierarchical structure by \citet{yao2023hierarchical}.
Some other research includes modeling an intersection controlled by traffic lights~\citep{loos2011safe};
using the State-Clock Logic for the representation of time~\citep{bischopink2022spatial};
and employing ontology~\citep{bagschik2018ontology,li2020ontology}.

The above methods for describing spatial relationships rely on either road segments or a grid-based occupancy approach, which have their limitations.
Distinguishing between consecutive free grids or road segments is usually not necessary,
and it can be crucial to describe the positional relationships of vehicles within the same grid or road segment. 
In this regard, a simple discretized description of the road network is often insufficient for test scenario generation.

\section{Highway Traffic Scenario Logic}
\label{sec:HTSL}
\subsection{Abstract Model}
\label{subsec:HTSL-syntax-semantics}

This section formally defines a scenario on a uni-direction multi-lane highway, where one or more vehicles travel on a road with multiple lanes.
The road length is infinite, and the number of lanes remains constant throughout.
Please note that a road with variable lane numbers can be modeled as separate roads, which will be discussed in \autoref{sec:UTSL}.
Vehicles can perform lane-changing maneuvers between lanes to switch to adjacent lanes.

We denote the set of vehicles as $\mathbb{C} = \{c_1, c_2, \dots, c_{N} \}$, where $N$ is the number of vehicles in the scenario.
The set of lanes is denoted as $\mathbb{L} = \{l_1, l_2, \dots, l_{M} \}$, where $M$ is the number of lanes.
Then, we can define a road as follows:
\begin{definition}[Road]
A road is an ordered arrangement of lanes of the lane set $\mathbb{L}$, denoted as $\Road=\langle l_{i_1},l_{i_2},\dots,l_{i_{M}}\rangle$.
A binary relation $\prec_{\text{left}}$ is defined on it, where $l_{i_j} \prec_{\text{left}} l_{i_{j+1}}, 1\leq j \leq M-1$.
\end{definition}

The lateral position of a vehicle (perpendicular to the direction of the road) is described using the lanes it occupies, while the longitudinal position (along the direction of the road) can be described using the positional relationship with other vehicles.
Using the road coordinate system, it can be defined as follows:
\begin{definition}[Longitudinal Positional Relationship]
\label{def:lonr}
Denote the range occupied by vehicle $c$ as $[s_c^-,s_c^+]$, where $s_c^-$ and $s_c^+$ represent the $s$-coordinates of the rear and front ends of the vehicle under Frenet coordinate system~\citep{werling2010optimal}, respectively.
Assuming that the $S$ axis direction aligns with the direction of the vehicle, we can categorize the longitudinal positional relationship between two vehicles $c_1$ and $c_2$ on the same road as follows:
\begin{enumerate}
    \item \textbf{Ahead}: If $ s_{c_1}^- > s_{c_2}^+ $, then $ c_1 $ is ahead $ c_2 $.
    \item \textbf{Cover}: If $ s_{c_1}^- \leq s_{c_2}^+ $ and $ s_{c_1}^+ \geq s_{c_2}^- $, then $ c_1 $ covers $ c_2 $.
    \item \textbf{Behind}: If $ s_{c_1}^+ < s_{c_2}^- $, then $ c_1 $ is behind $ c_2 $.
\end{enumerate}
The definition is similar when the $S$ axis direction is opposite to that of the vehicles.
Obviously, their longitudinal positional relationship falls into one of these three categories for two vehicles on the same road.
\end{definition}

Then we can define a \textit{scene}, which is a ``frame'' of a scenario:
\begin{definition}[Scene on Highway]
A scene on a uni-direction multi-lane highway is a pair $\Sc=(\mathscr{L},\mathscr{D})$, where:
\begin{itemize}
    \item[-] $\mathscr{L}: \mathbb{C}\mapsto \mathcal{P}(\mathbb{L})$ is a mapping from the set of vehicles $\mathbb{C}$ to the power set of the set of lanes $\mathcal{P}(\mathbb{L})$, assigning to each vehicle the set of lanes to which it belongs; note that vehicles may cross lane dividers, so a vehicle may belong to more than one lane.
    \item[-] $\mathscr{D}: \mathbb{C}\times\mathbb{C} \mapsto \mathbb{D}$ is a mapping from the set of vehicle pairs to the set of longitudinal positional relationship $\mathbb{D}=\{\Lahead,\Lcover,\Lbehind\}$.
\end{itemize}
\end{definition}

A scenario is a sequence of scenes, defined as follows:
\begin{definition}[Scenario on Highway]
A scenario on a uni-direction multi-lane highway is a quintuple $\LS=(\mathbb{C},\mathbb{L},\mathbb{D},T,\langle\Sc_0,\Sc_1,\dots,\Sc_{T-1}\rangle)$, where:
\begin{itemize}
    \item[-] The meaning of $\mathbb{C}$, $\mathbb{L}$ and $\mathbb{D}$ is as stated above;
    \item[-] $T\geq 1$ is the number of scenes included in the scenario, i.e. the length of the scenario;
    \item[-] $\Sc_i=(\mathscr{L}_i,\mathscr{D}_i)$ is the $i$-th scene that occurs in the scenario in time sequence.
\end{itemize}
\end{definition}

For convenience, further define the $i$-tail of a scenario as:
\begin{definition}[$i$-tail of a Scenario]
\label{def:tail-of-scenario}
The $i$-tail of scenario $\LS$, denoted by $\LS_{i:}$, is defined as $(\mathbb{C},\mathbb{L},\mathbb{D},T-i,\langle\Sc_i,\Sc_{i+1},\dots,\Sc_{T-1}\rangle)$ for all $0 \leq i < T$
\end{definition}

Now we provide the syntax and semantics of the TSL for uni-direction multi-lane highway scenario.

\begin{definition}[Syntax of Highway TSL]
\label{def:HTSL-syntax}
A formula of highway TSL can be recursively defined as:
\begin{equation*}
\begin{gathered}
\phi ::= \top \mid \Lon{c}{l} \mid \Lleft{l_1}{l_2} 
\mid \Llonr{c_1}{c_2}{d} \mid c_1 = c_2\\
\mid l_1 = l_2 \mid d_1 = d_2 
\mid \lnot \phi \mid \phi_1 \to \phi_2 \mid \Next \phi \mid \Globally \phi\\
\mid \Final \mid \forall c\, \phi \mid \forall l\, \phi \mid \forall d\, \phi 
\end{gathered}
\end{equation*}
where, $c,c_1,c_2\in\mathbb{C}$, $l,l_1,l_2\in\mathbb{L}$, $d,d_1,d_2\in\mathbb{D}$.
The modal operators $\Next$ and $\Globally$ take their meanings ``Next'' and ``Globally'' in temporal logic, respectively.
$\Final$ denotes that the length of this scenario is 1, indicating the final state.
\begin{remark}
We can further define $\phi\land \psi$ , $\phi\lor \psi$ , $\exists x\, \phi$ and $\Finally \phi$ as
$\lnot(\phi \to (\lnot \psi))$ , $(\lnot \phi) \to \psi$ , $\lnot \forall x\, (\lnot \phi)$ and $\lnot \Globally \lnot \phi$, respectively.
\end{remark}
\end{definition}

Please note that the temporal modal operators here represent the transition of scenes, indicating only the chronological order, not the exact time.
The interval between steps can last for an arbitrary duration and can be different.

\begin{definition}[Semantics of Highway TSL]
Given a road $\Road$ and a scenario $\LS$, the satisfaction of formulae in the \autoref{def:HTSL-syntax} follows the semantics listed below:
\begin{align*}
    &\Road,\,\LS \models \top && & &\text{ for all } \Road,\, \LS\\
    &\Road,\,\LS \models \Lon{c}{l} &\Leftrightarrow & & & l \in \mathscr{L}_0(c)\\
    &\Road,\,\LS \models \Lleft{l_1}{l_2} &\Leftrightarrow & & & l_1 \prec_{\text{left}} l_2\\
    &\Road,\,\LS \models \Llonr{c_1}{c_2}{d} &\Leftrightarrow & & & \mathscr{D}_0(c_1,c_2) = d\\
    &\Road,\,\LS \models \lnot \phi &\Leftrightarrow & & & \Road,\,\LS \nmodels \phi\\
    &\Road,\,\LS \models \phi_1 \to \phi_2 &\Leftrightarrow & & & \Road,\,\LS \models \phi_2 \;\text{if}\; \Road,\,\LS \models \phi_1 \\
    &\Road,\,\LS \models \Next \phi &\Leftrightarrow & & & T>1\;\text{and}\;\Road,\,\LS_{1:} \models \phi \\
    &\Road,\,\LS \models \Globally \phi &\Leftrightarrow & & & \forall 0 \leq i < T,\,\Road,\,\LS_{i:} \models \phi \\
    &\Road,\,\LS \models \Final &\Leftrightarrow & & & T=1
\end{align*}
\end{definition}

\subsection{Rules of Highway TSL}
\label{subsec:HTSL-constraints}

Realistic scenarios obey both physics rules and traffic rules.
In this subsection, we outline these rules and auxiliary predicates we use in the form of TSL formulae, as shown in \autoref{tab:htsl-rule}.
Traffic rules can be added or subtracted according to the laws of different regions.

\begin{table*}[t!]
\fontsize{9pt}{10pt}\selectfont
\centering
    \begin{tabular}{ll}
        \toprule
        Rule & TSL Formula \\
        \midrule
        PR1 & $\Globally (\forall c_1 \forall c_2\, \Llonr{c_1}{c_2}{\Lahead}
\to \Llonr{c_2}{c_1}{\Lbehind})$, \\
               & $\Globally (\forall c_1 \forall c_2\, \Llonr{c_1}{c_2}{\Lbehind} \to \Llonr{c_2}{c_1}{\Lahead})$,                     \\
               & $\Globally (\forall c_1 \forall c_2\, \Llonr{c_1}{c_2}{\Lcover} \to \Llonr{c_2}{c_1}{\Lcover})$.\\
        \addlinespace
        PR2 & $\Globally ( \forall c_1 \forall c_2 \forall c_3 \, \Llonr{c_1}{c_2}{\Lahead} 
\land\Llonr{c_2}{c_3}{\Lahead} \to\Llonr{c_1}{c_3}{\Lahead})$, \\
                   & $\Globally ( \forall c_1 \forall c_2 \forall c_3 \, \Llonr{c_1}{c_2}{\Lbehind} 
\land\Llonr{c_2}{c_3}{\Lbehind} \to\Llonr{c_1}{c_3}{\Lbehind})$.\\
        \addlinespace
        PR3 & $\Globally (\forall c_1 \forall c_2 \forall c_3 \, \lnot(\Llonr{c_1}{c_2}{\Lahead} \land\Llonr{c_2}{c_3}{\Lcover} \land\Llonr{c_3}{c_1}{\Lahead}))$. \\
        \addlinespace
        PR4 & $\Final \lor \Globally (\forall c_1 \forall c_2\, \Llonr{c_1}{c_2}{\Lahead} 
\to \Next (\Llonr{c_1}{c_2}{\Lahead} \lor \Llonr{c_1}{c_2}{\Lcover}))$, \\
                 & $\Final \lor \Globally (\forall c_1 \forall c_2\, \Llonr{c_1}{c_2}{\Lbehind} 
\to \Next (\Llonr{c_1}{c_2}{\Lbehind} \lor \Llonr{c_1}{c_2}{\Lcover}))$. \\
        \addlinespace
        PR5 & $\Globally ( \forall c \forall l_1 \forall l_2 \forall l_3 \,\Lon{c}{l_1} \land \Lon{c}{l_2} \land \Lcleft{l_1}{l_3} \land \Lcleft{l_3}{l_2} \to \Lon{c}{l_3})$. \\
        \addlinespace
        PR6 & $\Globally (\forall c \exists l\, \Lon{c}{l})$. \\
        \addlinespace
        PR7 & $\Final \lor \Globally(\forall c \forall l_1 \forall l_2 \,\lnot(\Lon{c}{l_1} \land \Lon{c}{l_2} \land \lnot(l_1 = l_2) \land \lnot \Next \Lon{c}{l_1} \land  \lnot \Next \Lon{c}{l_2}))
$, \\
               & $\Final \lor \Globally(\forall c \forall l_1 \forall l_2 \,\lnot(\lnot\Lon{c}{l_1} \land \lnot\Lon{c}{l_2} \land \lnot(l_1 = l_2) \land \Next \Lon{c}{l_1} \land  \Next \Lon{c}{l_2}))$, \\
               & $\Final \lor \Globally(\forall c \forall l_1 \forall l_2 \,\lnot(\Lon{c}{l_1} \land \lnot\Lon{c}{l_2} \land \lnot(l_1 = l_2) \land \lnot \Next \Lon{c}{l_1} \land  \Next \Lon{c}{l_2}))$.\\
        \addlinespace
        TR1 &$\Globally (\forall c \forall l_1 \forall l_2 \forall l_3 \,\lnot (\Lon{c}{l_1} \land \Lon{c}{l_2} \land \Lon{c}{l_3}  \land \lnot(l_1 = l_2)\land \lnot(l_1 = l_3)\land \lnot(l_2 = l_3) ))$.  \\
        \addlinespace
        TR2 & $\Globally (\forall c_1 \forall c_2 \forall l \, \lnot (\Lon{c_1}{l} \land \Lon{c_2}{l} \land \Llonr{c_1}{c_2}{\Lcover} \land \lnot(c_1 = c_2) ))$. \\
        \addlinespace
        AP1 & $\forall l_1 \forall l_2 \, \Lleft{l_1}{l_2} \to \Lcleft{l_1}{l_2}$, \\
        & $\forall l_1 \forall l_2 \forall l_3 \,\Lleft{l_2}{l_3} \land \Lcleft{l_1}{l_2} \to \Lcleft{l_1}{l_3}$. \\
        \bottomrule
    \end{tabular}
    \caption{The rules and auxiliary predicates used in Highway TSL}
    \label{tab:htsl-rule}
\end{table*}

\textbf{Physics Rules}

\begin{enumerate}[labelindent=0em,labelwidth=2em,itemindent=0em,leftmargin=!]
\item[PR1] The longitudinal positional relationships are symmetric.

\item[PR2] The longitudinal positional relationships are transitive.

\item[PR3] Longitudinal positional relationships must not be contradictory.


\item[PR4] The longitudinal positional relationships must change continuously.

\item[PR5] The occupied lanes by a vehicle must be one or more adjacent lanes.
We utilize an auxiliary predicate $\atom{cleft/2}$ (AP1) here for simplicity, representing one lane being to the left of another lane but not necessarily adjacent.

\item[PR6] Each vehicle always occupies at least one lane.

\item[PR7] The occupied lanes must change continuously.
That is, only one lane can be added to or removed from the set of occupied lanes on each step.

\end{enumerate}

\textbf{Traffic Rules}

\begin{enumerate}[labelindent=0em,labelwidth=2em,itemindent=0em,leftmargin=!]
\item[TR1] Vehicles are not allowed to occupy three or more lanes simultaneously.

\item[TR2] Two vehicles cannot drive side by side in the same lane.
\end{enumerate}

A realistic scenario and its tails (because a tail of a realistic scenario is also a realistic scenario) must satisfy all the aforementioned rules.
Consequently, by utilizing these rules, we can validate the plausibility of a scenario, rendering the reasoning of scenarios under TSL executable.

\section{Urban Traffic Scenario Logic}
\label{sec:UTSL}
In this section, we will introduce the extension of TSL to general urban traffic scenarios, including vehicles traveling on structured roads without pedestrians.


\subsection{Formal Representation of Urban Road Network}
\label{subsec:urban-road-network}
\begin{figure}[t]
    \centering
    \hfill
    \subcaptionbox{Intersection\label{fig:cross-map}}{\includegraphics[height=0.9in]{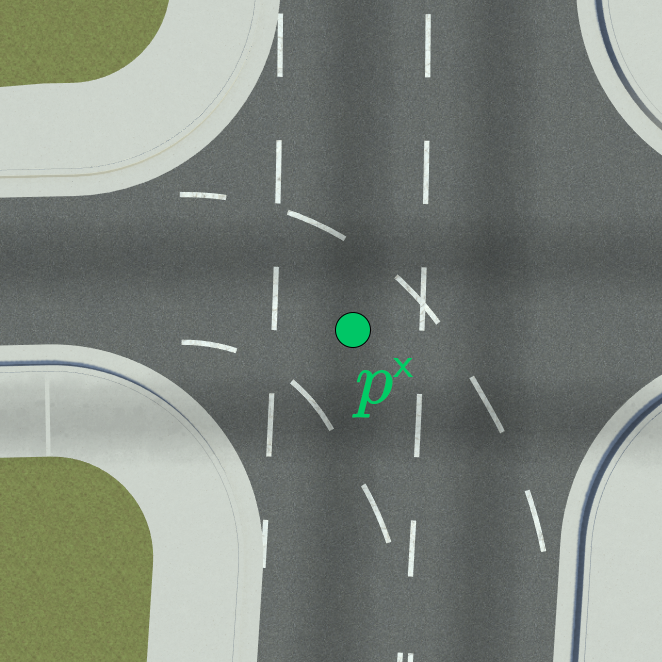}}
    \hfill
    \subcaptionbox{Connection\label{fig:connect-map}}{\includegraphics[height=0.9in]{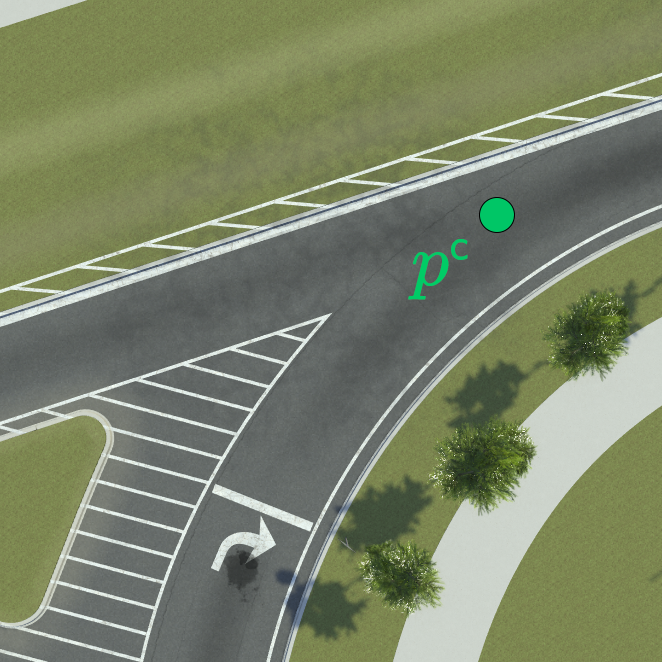}}
    \hfill
    \subcaptionbox{Overlap\label{fig:overlap-map}}{\includegraphics[height=0.9in]{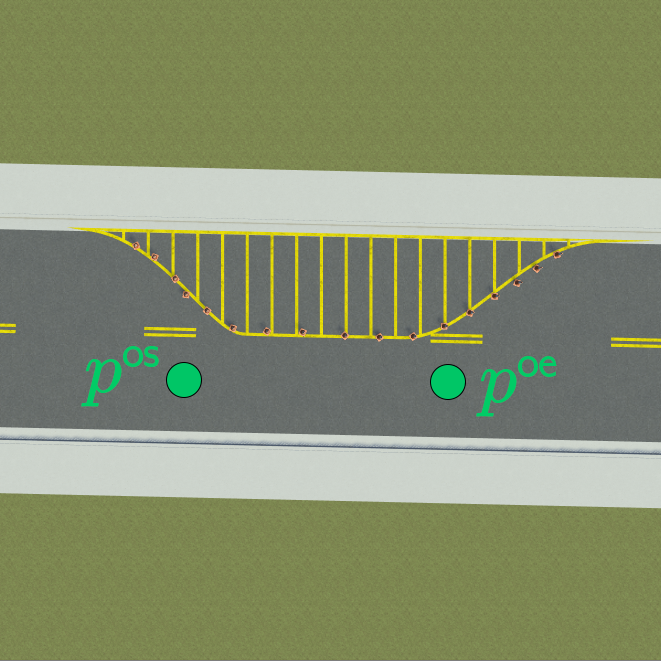}}
    \hfill
    \caption{Spectial points on urban road network}
\end{figure}

A structured urban road network consists of lanes.
When considering two lanes, their relationships can be categorized as follows:

\begin{enumerate}
    \item They belong to the same road, in which case the scenario is identical to that described in \autoref{sec:HTSL};
    \item They have an intersection point, but are not connected at this point (\autoref{fig:cross-map});
    \item They are connected at a connection point, where vehicles can switch to another lane (\autoref{fig:connect-map});
    \item They have an overlap segment in which vehicles travel in the same direction. In this case, we can merge the overlap segment into a single lane;
    \item They have an overlap segment in which vehicles travel in opposite directions (\autoref{fig:overlap-map});
    \item They are unrelated.
\end{enumerate}
All we need is to extend the TSL for intersections, connections, and overlaps in opposite directions.
The rest of the cases can reuse Highway TSL.


\begin{enumerate}
\item Intersection\\
Without loss of generality, we only consider the situation where two lanes intersect once.
For multiple intersections, we can simply apply it repeatedly to each intersection.

Denote the intersection point as $p^\mathsf{x}$. Then, for any vehicle $c$ on these two intersecting lanes, there are essentially three positional relationships with respect to this intersection point: yet to pass, currently passing, and already passed.
We use the predicate $\atom{lonpr/3}$ to represent these relationships, corresponding to the predicate $\atom{lonr/3}$ for vehicles' relationships.
The same predicate can also be used for the connection points and endpoints of overlap segments below.

\item Connection\\
Two or more lanes can connect at their endpoints.
We denote the connection points as $p^\mathsf{c}$. Depending on the direction of travel of vehicles on the lanes, there exists a successor relationship $\succ_\mathsf{c}$ between the connection points and the lanes.

\item Overlap in opposite directions\\
When examining the overlap segment between points $p^\mathsf{os}$ and $p^\mathsf{oe}$, as shown in \autoref{fig:overlap-map}, we can designate one of the two overlapping lanes' direction as the reference direction, i.e., from $p^\mathsf{os}$ to $p^\mathsf{oe}$.
Another longitudinal positional relationship, denoted as $\atom{lonro/3}$, exists between the vehicles in the overlap segment, and:
\begin{itemize}
    \item For the lane in the reference direction, the relationships $\atom{lonro/3}$ between vehicles are the same as $\atom{lonr/3}$;
    \item For the lane in the opposite direction, the relationships $\atom{lonro/3}$ between vehicles are opposite to $\atom{lonr/3}$;
    \item For two vehicles belonging to different directions, the relationship $\atom{lonro/3}$ between them can be arbitrary, but only monotonic changes are allowed.
\end{itemize}
\end{enumerate}

Besides, when there are multiple points (including intersection points, connection points, and endpoints of overlap segments) on a lane $l$, there also exists a successor relationship $\succ_\mathsf{p}^l$ between points according to the direction of vehicle travel on the lane.

Based on the above analysis, we further provide a formal definition of the road network:

\begin{definition}[Road Network]
\label{def:roadnetwork}
The road network is a nonuple $\Map = (\mathbb{R},\mathbb{P}^\mathsf{X},\mathbb{P}^\mathsf{C},\mathbb{P}^\mathsf{OS},\mathbb{P}^\mathsf{OE},\mathbb{S}^\mathsf{P},\mathbb{S}^\mathsf{C}, \mathbb{O},\mathbb{T})$, where:
\begin{itemize}
    \item[-] $\mathbb{R}=\{\mathcal{R}_1,\mathcal{R}_2,\dots\}$ is the set of all roads;
    \item[-] $\mathbb{P}^\mathsf{X},\mathbb{P}^\mathsf{C},\mathbb{P}^\mathsf{OS},\mathbb{P}^\mathsf{OE}$are sets of points $p^\mathsf{x}$, $p^\mathsf{c}$, $p^\mathsf{os}$, and $p^\mathsf{oe}$ respectively;
    \item[-] 
    $\mathbb{S}^\mathsf{P} \subseteq \mathbb{L}\times\mathbb{P}\times\mathbb{P}$ defines the successor relationship $\succ_\mathsf{p}^l$ for points on each lane $l$, where $\mathbb{L}=\bigcup_i \mathbb{L}_i$ is the set of all lanes, and $\mathbb{P}=\mathbb{P}^\mathsf{X}\cup\mathbb{P}^\mathsf{C}\cup\mathbb{P}^\mathsf{OS}\cup\mathbb{P}^\mathsf{OE}$ is the set of all points;
    \item[-] $\mathbb{S}^\mathsf{C} \subseteq \mathbb{P}^\mathsf{C}\times\mathbb{L}$ defines the successor lane for connection points, denoted as the relation $\succ_\mathsf{c}$;
    \item[-] $\mathbb{O} \subseteq \mathbb{P}^\mathsf{OS}\times\mathbb{P}^\mathsf{OE}$ represents the pairs $(p^\mathsf{os},p^\mathsf{oe})$ for each overlap segment.
    \item[-] $\mathbb{T} \subseteq \mathbb{L}\times\mathbb{P}$ defines the affiliation relationship between points and lanes.
\end{itemize}
\end{definition}
We proposed an example of this abstraction in the Technical Appendix.

\begin{theorem}
For a structured 2D OpenDRIVE road network, there exists a representation in the form defined by \autoref{def:roadnetwork}.
\begin{proof}
See the discussion above.
\end{proof}
\end{theorem}

\subsection{Abstract Model}

In this subsection, we provide the syntax and semantics of TSL for urban traffic scenarios.

\begin{definition}[Scene]
\label{def:urban-scene}
A (urban) scene is a quadruple $\Sc=(\mathscr{L},\mathscr{D},\mathscr{P},\mathscr{Q})$, where:
\begin{itemize}
    \item[-] $\mathscr{L}: \mathbb{C}\mapsto \mathcal{P}(\mathbb{L})$ is a mapping from the set of vehicles to the power set of the set of lanes, assigning to each vehicle a set of lanes to which it occupies;
    \item[-] $\mathscr{D}: \mathbb{C}\times\mathbb{C} \mapsto \mathbb{D}$ is a mapping from pairs of vehicles to the set of longitudinal positional relationships $\mathbb{D}=\{\Lahead,\Lcover,\Lbehind,\Lnone\}$, where $\Lnone$ indicates that there is no relationship between two vehicles, which occurs when the two vehicles belong to different roads;
    \item[-] $\mathscr{P}: \mathbb{C}\times\mathbb{P} \mapsto \mathbb{D}$ is a mapping from pairs of vehicles and points to the set of longitudinal positional relationships, as explained in \autoref{subsec:urban-road-network};
    \item[-] $\mathscr{Q}: \mathbb{C}\times\mathbb{C} \mapsto \mathbb{D}$ is a mapping to define the temporary longitudinal positional relationship $\atom{lonro/3}$ on the overlap segments.
\end{itemize}
\end{definition}

\begin{definition}[Scenario]
\label{def:urban-scenario}
A (urban) scenario is a sextuple $\LS=(\mathbb{C},\mathbb{L},\mathbb{D},\mathbb{P},T,\langle\Sc_0,\Sc_1,\dots,\Sc_{T-1}\rangle)$, where:
\begin{itemize}
    \item[-] $\mathbb{C}$, $\mathbb{L}$, $\mathbb{D}$, and $\mathbb{P}$ have the meanings as described above;
    \item[-] $T\geq 1$ is the length of the scenario;
    \item[-] $\Sc_i=(\mathscr{L}_i,\mathscr{D}_i,\mathscr{P}_i,\mathscr{Q}_i)$ is the $i$-th scene that occurs in the scenario in time sequence.
\end{itemize}
\end{definition}
The definition of the $i$-tail of a scenario is similar to \autoref{def:tail-of-scenario} and will not be repeated.

\begin{theorem}
\label{thm:scenario-existance}
For a pedestrian-free traffic scenario on road network $\Map$, there exists a representation in the form defined by \autoref{def:urban-scenario}.
\begin{proof}
Since the longitudinal relationship as defined by \autoref{def:lonr} encompasses all situations,
the mappings $\mathscr{L}$, $\mathscr{D}$, $\mathscr{P}$, and $\mathscr{Q}$ are guaranteed to exist and be unique for each scene.
Consequently, we establish the existence of a representation $\LS$ for each pedestrian-free traffic scenario.
\end{proof}
\end{theorem}

\begin{definition}[Syntax of Urban TSL]
\label{def:UTSL-syntax}
A formula of urban TSL can be recursively defined as follows:
\begin{equation*}
\begin{gathered}
\phi ::= \top \mid \Lon{c}{l} \mid \Lpon{p}{l} \mid \Lleft{l_1}{l_2}
\mid \Llonr{c_1}{c_2}{d}\\
\mid \Llonpr{c}{p}{d} \mid \Llonro{c_1}{c_2}{d} \mid \Lsuccp{l}{p_1}{p_2} \\
\mid \Lsuccl{p}{l} \mid \Lbelong{l}{\Road} \mid \Loverlap{p_1}{p_2} \mid \Lispx{p} \mid \Lispc{p} \\
\mid \Lispos{p} \mid \Lispoe{p} \mid c_1 = c_2 \mid l_1 = l_2 \mid d_1 = d_2 \mid p_1 = p_2 \\
\mid \Road_1 = \Road_2 \mid \lnot \phi \mid \phi_1 \to \phi_2 \mid \Next \phi \mid \Globally \phi \mid \Final \\
\mid \forall c\, \phi \mid \forall l\, \phi \mid \forall d\, \phi \mid \forall p\, \phi \mid \forall \Road\, \phi
\end{gathered}
\end{equation*}
\end{definition}

\begin{definition}[The Semantics of Urban TSL]
Given the road network $\Map$ and the logical scenario $\LS$, the satisfaction of formulas defined in \autoref{def:UTSL-syntax} follows the following semantics:
\begingroup
\allowdisplaybreaks
\begin{align*}
    &\Map,\,\LS \models \top && & &\text{ for all } \Map,\, \LS\\
    &\Map,\,\LS \models \Lon{c}{l} &\Leftrightarrow & & & l \in \mathscr{L}_0(c)\\
    &\Map,\,\LS \models \Lpon{p}{l} &\Leftrightarrow & & & (l,p) \in \mathbb{T}\\
    &\Map,\,\LS \models \Lleft{l_1}{l_2} &\Leftrightarrow & & & l_1 \prec_{\text{left}} l_2\\
    &\Map,\,\LS \models \Llonr{c_1}{c_2}{d} &\Leftrightarrow & & & \mathscr{D}_0(c_1,c_2) = d\\
    &\Map,\,\LS \models \Llonpr{c}{p}{d} &\Leftrightarrow & & & \mathscr{P}_0(c,p) = d\\
    &\Map,\,\LS \models \Llonro{c_1}{c_2}{d} &\Leftrightarrow & & & \mathscr{Q}_0(c_1,c_2) = d\\
    &\Map,\,\LS \models \Lsuccp{l}{p_1}{p_2} &\Leftrightarrow & & & (l,p_1,p_2)\in\mathbb{S}^\mathsf{P}\\
    &\Map,\,\LS \models \Lsuccl{p}{l} &\Leftrightarrow & & & (p,l)\in\mathbb{S}^\mathsf{C}\\
    &\Map,\,\LS \models \Lbelong{l}{\Road} &\Leftrightarrow & & & l\in \Road\\
    &\Map,\,\LS \models \Loverlap{p_1}{p_2} &\Leftrightarrow & & & (p_1,p_2)\in\mathbb{O}\\
    &\Map,\,\LS \models \Lispx{p} &\Leftrightarrow & & & p \in \mathbb{P}^\mathsf{X}\\
    &\Map,\,\LS \models \Lispc{p} &\Leftrightarrow & & & p \in \mathbb{P}^\mathsf{C}\\
    &\Map,\,\LS \models \Lispos{p} &\Leftrightarrow & & & p \in \mathbb{P}^\mathsf{OS}\\
    &\Map,\,\LS \models \Lispoe{p} &\Leftrightarrow & & & p \in \mathbb{P}^\mathsf{OE}\\
    &\Map,\,\LS \models \lnot \phi &\Leftrightarrow & & & \Map,\,\LS \nmodels \phi\\
    &\Map,\,\LS \models \phi_1 \to \phi_2 &\Leftrightarrow & & & \Map,\LS \models \phi_2 \;\text{if}\; \Map,\LS \models \phi_1 \\
    &\Map,\,\LS \models \Next \phi &\Leftrightarrow & & & T>1\;\text{and}\;\Map,\,\LS_{1:} \models \phi \\
    &\Map,\,\LS \models \Globally \phi &\Leftrightarrow & & & \forall 0 \leq i < T,\,\Map,\,\LS_{i:} \models \phi \\
    &\Map,\,\LS \models \Final &\Leftrightarrow & & & T=1
\end{align*}%
\endgroup
\end{definition}

\begin{table*}[t!]
\fontsize{9pt}{10pt}\selectfont
\setlength{\tabcolsep}{1mm}
\centering
    \begin{tabular}{ll}
        \toprule
        Rule & TSL Formula \\
        \midrule
        PR8 & $\Globally (\forall c \, \forall \Road_1 \forall \Road_2 \, \lnot  (\Lcbelong{c}{\Road_1} \land \Lcbelong{c}{\Road_2} \land \lnot(\Road_1 = \Road_2)))$. \\
        \addlinespace
        PR9 & $\Final \lor \Globally(\forall c \forall p \, \Llonpr{c}{p}{\Lbehind} \to \Next (\Llonpr{c}{p}{\Lbehind}\lor \Llonpr{c}{p}{\Lcover} \lor \Llonpr{c}{p}{\Lnone}))$, \\
                   & $\Final \lor \Globally(\forall c \forall p \, \Llonpr{c}{p}{\Lcover} \to \Next (\Llonpr{c}{p}{\Lcover}\lor \Llonpr{c}{p}{\Lahead} \lor \Llonpr{c}{p}{\Lnone}))$,\\
                   & $\Final \lor \Globally(\forall c \forall p \, \Llonpr{c}{p}{\Lahead} \to
\Next (\Llonpr{c}{p}{\Lahead} \lor \Llonpr{c}{p}{\Lnone}))$.\\
        \addlinespace
        PR10 & $\Globally (\forall c \forall p \forall d_1 \forall d_2 \, \lnot (\Llonpr{c}{p}{d_1} 
\land \Llonpr{c}{p}{d_2} \land \lnot(d_1 = d_2)))$, \\
                   & $\Globally (\forall c \forall l \forall p \, \lnot (\Lon{c}{l} \land \Lpon{p}{l} \land \Llonpr{c}{p}{\Lnone} ))$. \\
        \addlinespace
        PR11 & $\Globally (\forall c_1 \forall c_2 \forall p \, \lnot (\Llonpr{c_1}{p}{\Lcover} 
\land \Llonpr{c_2}{p}{\Lcover} \land \lnot (c_1 = c_2)))$. \\
        \addlinespace
        PR12 & $\Final \lor \Globally (\forall c \forall p \forall l_1 \,\Llonpr{c}{p}{\Lcover} \land \Lispc{p} \land \Lon{c}{l_1} \to$ \\ & $\quad\quad \Next ((\Llonpr{c}{p}{\Lcover} \land \Lon{c}{l_1}) \lor \exists l_2(\Llonpr{c}{p}{\Lahead} \land \Lon{c}{l_2} \land \Lsuccl{p}{l_2})))$. \\
        \addlinespace
        PR13 & $\Globally (\forall c_1 \forall c_2 \forall p_1 \forall p_2 \forall d \, \Lfwdover{c_1}{p_1}{p_2}\land \Lfwdover{c_2}{p_1}{p_2} \land \Llonr{c_1}{c_2}{d} \to \Llonro{c_1}{c_2}{d})$, \\
                 & $\Globally (\forall c_1 \forall c_2 \forall p_1 \forall p_2 \, \Lrvsover{c_1}{p_1}{p_2} \land \Lrvsover{c_2}{p_1}{p_2} \land \Llonr{c_1}{c_2}{\Lahead} \to \Llonro{c_1}{c_2}{\Lbehind})$, \\
                 & $\Globally (\forall c_1 \forall c_2 \forall p_1 \forall p_2 \,\Lrvsover{c_1}{p_1}{p_2} \land \Lrvsover{c_2}{p_1}{p_2} \land \Llonr{c_1}{c_2}{\Lcover} \to \Llonro{c_1}{c_2}{\Lcover})$,\\
                 & $\Globally (\forall c_1 \forall c_2 \forall p_1 \forall p_2 \,\Lrvsover{c_1}{p_1}{p_2} \land \Lrvsover{c_2}{p_1}{p_2} \land \Llonr{c_1}{c_2}{\Lbehind} \to \Llonro{c_1}{c_2}{\Lahead})$, \\
                 & $\Globally (\forall c_1 \forall c_2 \forall p_1 \forall p_2 \,\Lfwdover{c_1}{p_1}{p_2} \land \Lrvsover{c_2}{p_1}{p_2} \to (\exists d\, \Llonro{c_1}{c_2}{d} \land \lnot (d = \Lnone)))$.\\
        \addlinespace
        AP2 & $\forall c \forall l \forall \Road \, \Lbelong{l}{\Road} \land \Lon{c}{l} \to \Lcbelong{c}{\Road}$. \\
        \addlinespace
        AP3 & $\forall c \forall p_1 \forall p_2 \forall l \, \Llonpr{c}{p_1}{\Lahead}\! \land\! \Llonpr{c}{p_2}{\Lbehind}\! \land\! \Lon{c}{l} \!\land\! \Lpon{p_1}{l} \!\land\! \Lpon{p_2}{l} \!\land\! \Loverlap{p_1}{p_2} \!\to\! \Lfwdover{c}{p_1}{p_2}$. \\
        \addlinespace
        AP4 & $\forall c \forall p_1 \forall p_2 \forall l \, \Llonpr{c}{p_1}{\Lbehind} \!\land\! \Llonpr{c}{p_2}{\Lahead} \!\land\! \Lon{c}{l}  \!\land\! \Lpon{p_1}{l} \!\land\! \Lpon{p_2}{l} \!\land\! \Loverlap{p_1}{p_2} \!\to\! \Lrvsover{c}{p_1}{p_2}$. \\
        \bottomrule
    \end{tabular}
    \caption{The additional rules and auxiliary predicates used in Urban TSL}
    \label{tab:utsl-rule}
\end{table*}

\subsection{Rules of Urban TSL}
\label{subsec:UTSL-constraints}
For urban traffic scenarios, the rules described in \autoref{subsec:HTSL-constraints} regarding vehicles within the same road still apply and will not be reiterated here.
Additionally, the following rules are added to handle situations across roads, whose formulae are shown in \autoref{tab:utsl-rule}.

\begin{enumerate}[labelindent=0em,labelwidth=2em,itemindent=0em,leftmargin=!]
\item[PR8] A vehicle can only be located on one road.
We introduce an auxiliary predicate $\atom{cbelong/2}$ (AP2) to represent the belonging relationship between a vehicle and a road.

\item[PR9] The longitudinal positional relationship between a vehicle and the points on the same lane can only change monotonically.

\item[PR10] A unique longitudinal relationship, which is not $\Lnone$, exists between a vehicle and the points on the lane it occupies.

\item[PR11] Only one vehicle is allowed to cover a point.

\item[PR12] When a vehicle passes through a connection point, it will enter one successor lane of that connection point.

\item[PR13] Rules for overlap segments following the discussion in \autoref{subsec:urban-road-network}.
Auxiliary predicates $\atom{fwdover/3}$ (AP3) and $\atom{rvsover/3}$ (AP4) are imported for this rule to indicate the vehicle driving along/against the reference direction on an overlap segment, respectively.

\item[PR14] All longitudinal positional relationships, including those between vehicles and vehicles, vehicles and points, and those on overlap segments, follow similar rules as presented in \autoref{subsec:HTSL-constraints}, which will not be reiterated here.
\end{enumerate}

By adding the above rules, we can validate the plausibility of an urban traffic scenario.

\section{Application}
\label{sec:application}

\begin{figure*}[t]
    \centering
    \subcaptionbox{Highway\label{fig:example-highway}}{\includegraphics[height=1.0in]{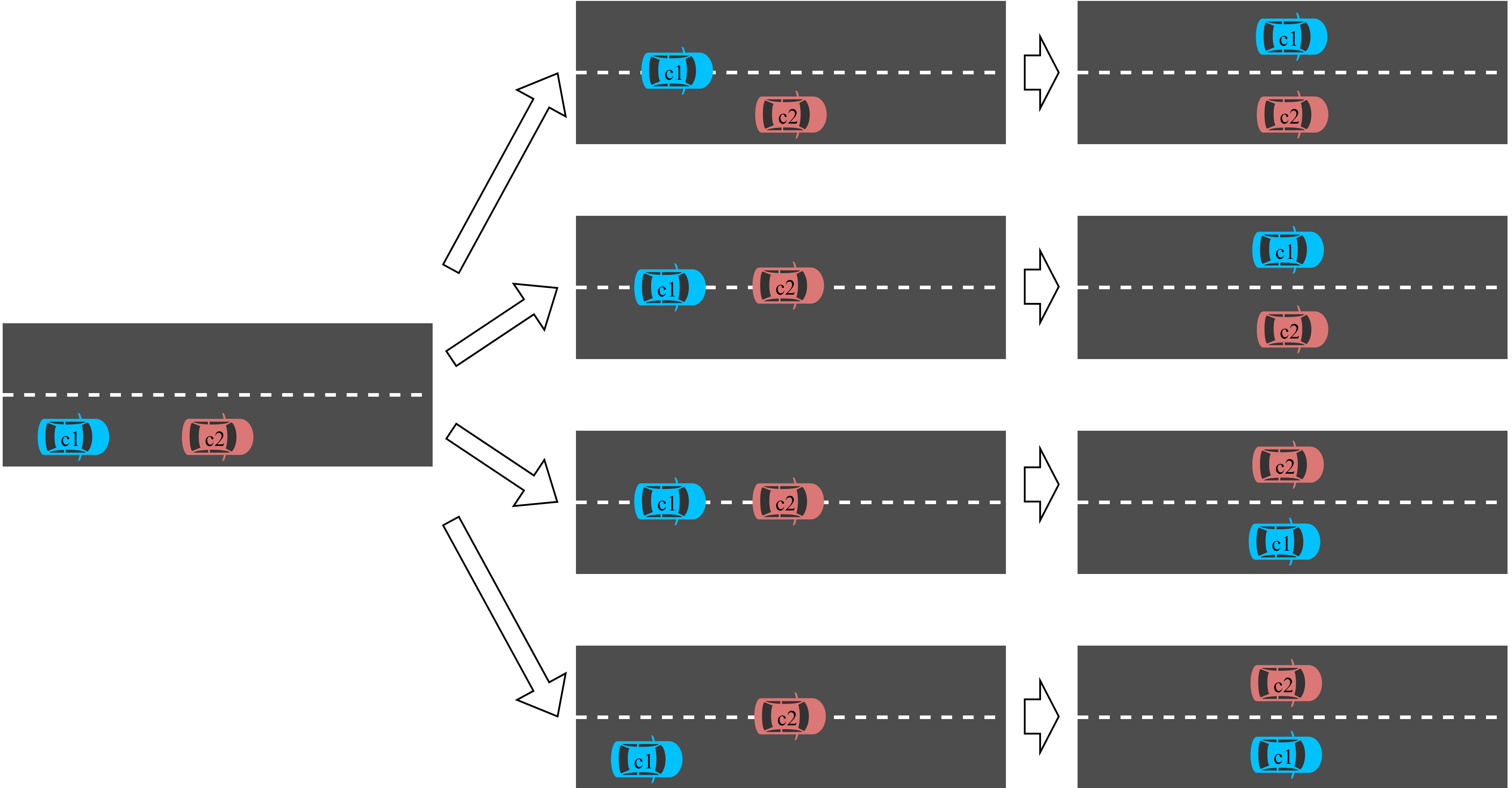}}
    \hfill
    \subcaptionbox{Intersection\label{fig:example-intersect}}{\includegraphics[height=1.0in]{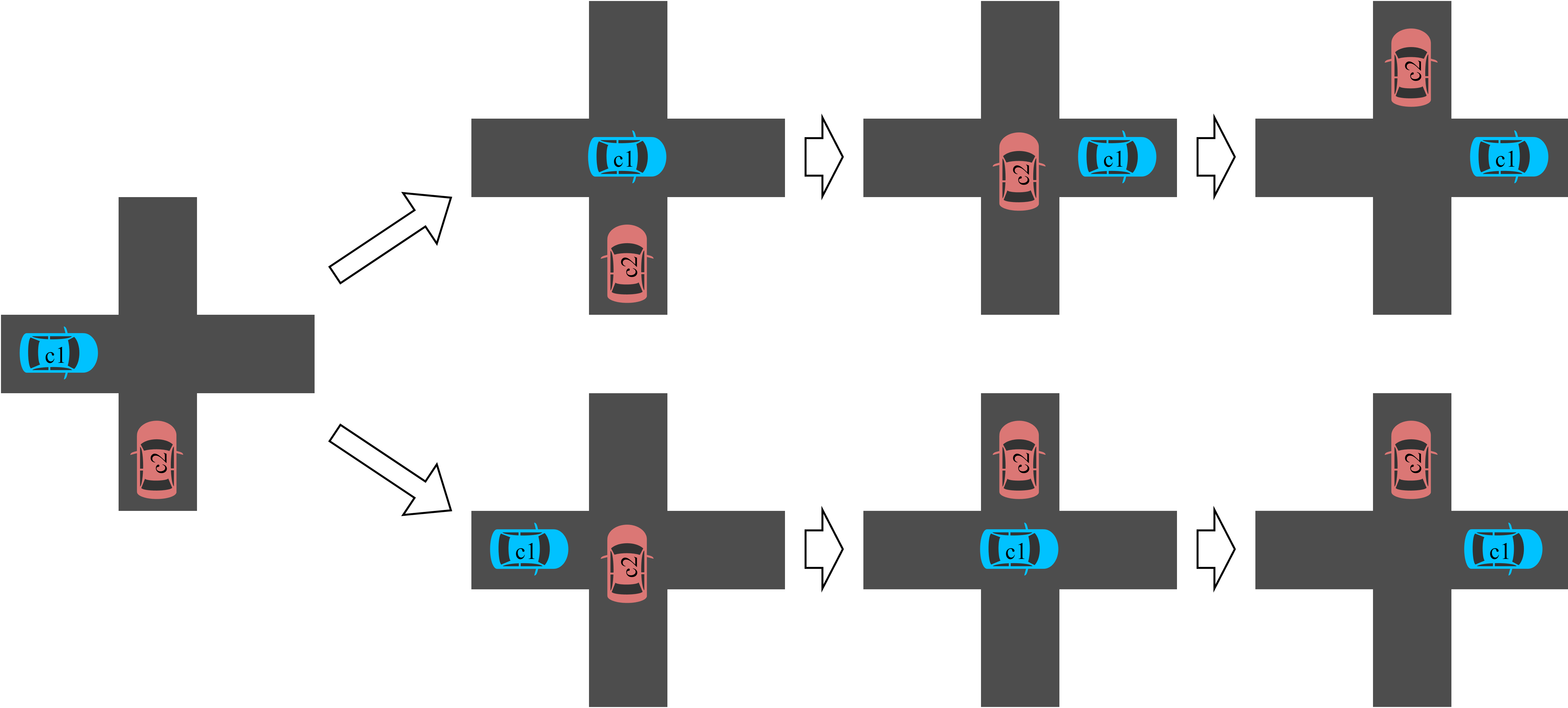}}
    \hfill
    \subcaptionbox{Connecting Roads\label{fig:example-connect}}{\includegraphics[height=1.0in]{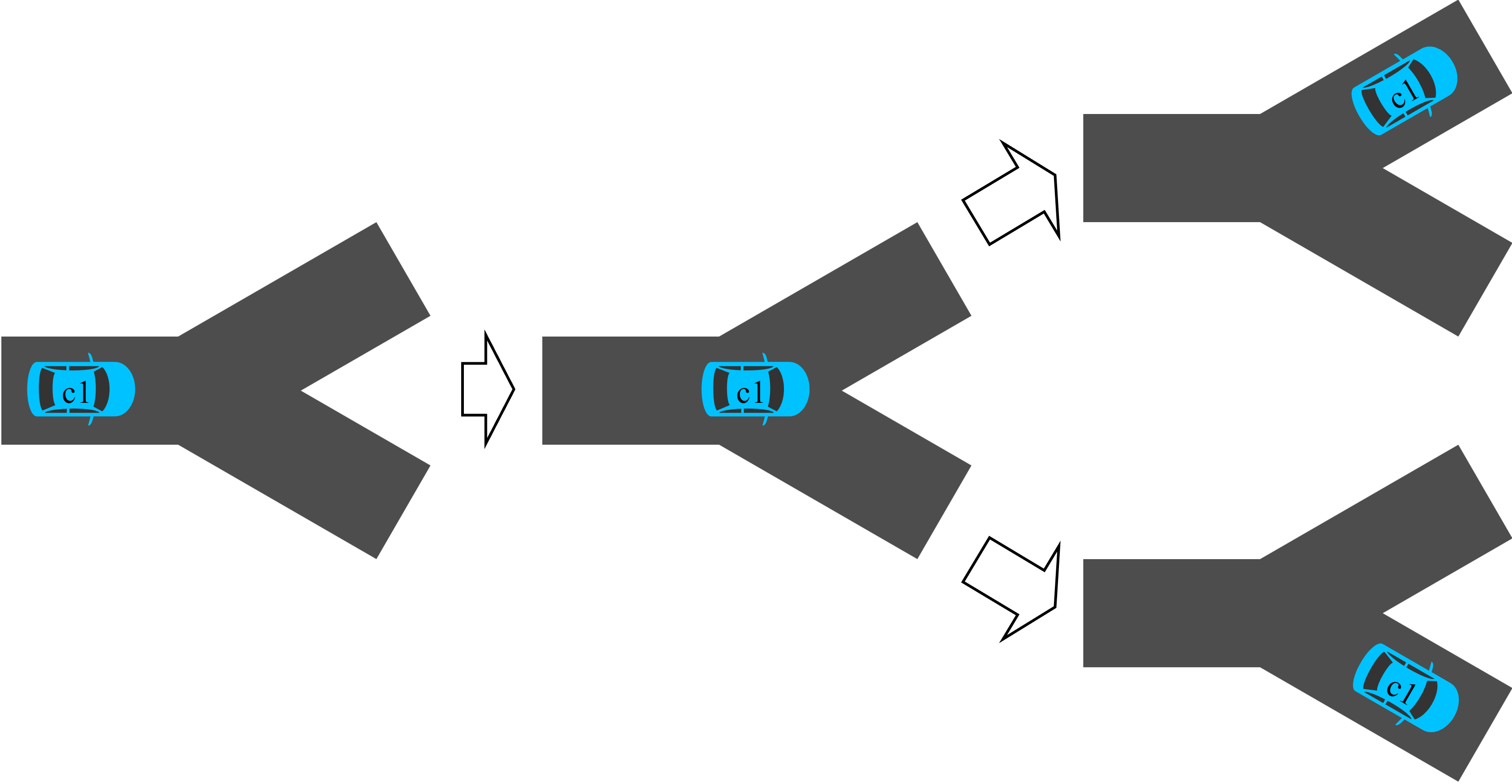}}
    \subcaptionbox{A vehicle passes two intersection points\label{fig:example-two-intersect}}{\includegraphics[width=0.8\textwidth]{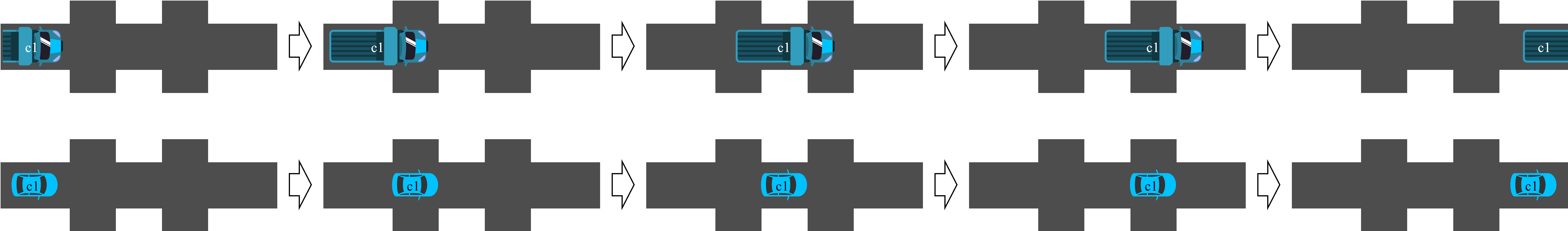}}
    \subcaptionbox{Overtake on the opposite side (overlapping lanes)\label{fig:example-overlap}}{\includegraphics[width=0.8\textwidth]{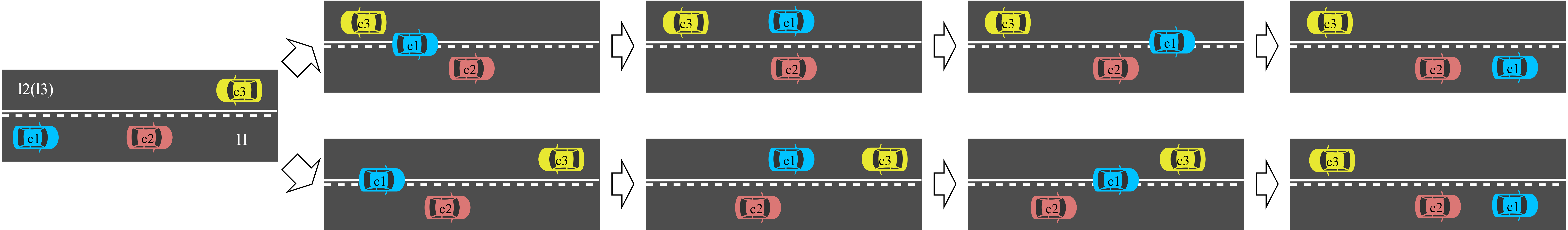}}
    \caption{Examples of generated scenarios with TSL}
    \label{fig:example}
\end{figure*}

TSL is a useful tool for modeling and reasoning about traffic scenarios.
It can generate various traffic scenarios, which can be used to create a library of test cases for comprehensive testing of autonomous vehicles.
Additionally, TSL can identify potential failure risks while developing autonomous driving strategies.
Its unique spatial representation makes it well-suited for these applications, which require both the comprehensiveness and generalization of scenarios.

In this section, we start by introducing the reasoning of TSL, which we implemented using Answer Set Programming (ASP) with temporal logic extensions.
Following that, we illustrate several examples of TSL for scenario generation to demonstrate its effectiveness in handling miscellaneous urban scenarios.

\subsection{Reasoning of TSL}

Given an initial scene $\Sc_0$, the set of vehicles $\mathbb{C}$, a finite horizon $T$ and a road network $\Map$, the reasoning of TSL is a kind of model expansion, i.e. to find all logical scenarios $\LS=(\mathbb{C},\mathbb{L},\mathbb{D},\mathbb{P},T,\langle\mathcal{S}_0,\mathcal{S}_1,\dots,\mathcal{S}_{T-1}\rangle)$ such that $\Map,\LS\models\Phi$, where $\Phi$ is the set of rules explained in \autoref{sec:HTSL} and \autoref{sec:UTSL}.

We use Telingo~\citep{cabalar2019telingo} to implement the reasoning of TSL.
Telingo is a solver for temporal logic programs based on ASP and the well-known Clingo~\citep{gebser2019multi} input language.
Telingo extends the ASP language by importing temporal modal operators under the semantics of the Linear Temporal Logic ($\mathcal{LTL}$)~\citep{cabalar2018temporal,pnueli1977temporal}.

We use aggregates~\citep{son2007constructive} instead of actions and transitions in the logic program we used for TSL reasoning to derive different scenarios.
All rules introduced in \autoref{sec:HTSL} and \autoref{sec:UTSL} are rewritten into corresponding ASP rules or constraints.
The road network and initial scene are described by facts.
For the entire implementation of our reasoning programs, please refer to our source code.

\subsection{Examples of Scenario Generation}

In this subsection, we provide several examples of TSL on different road layouts to provide an overview of TSL's features in scenario reasoning.

\begin{example}[Highway overtake]
\label{eg:highway}
\autoref{fig:example-highway} shows overtake scenarios on a uni-direction two-lane highway.
The shortest scenarios we get are shown in \autoref{fig:example-highway}.
In the first scenario, \texttt{c1} (the blue one) overtakes \texttt{c2} (the red one) by lane change.
In the second and third scenarios, the two cars change lanes simultaneously,
and then one of the cars gives up midway through the lane change.
In the fourth scenario, \texttt{c2} changes lanes to the left to make way for \texttt{c1}.
\end{example}

\begin{example}[Intersection]
When two cars pass an intersection point, TSL gives two different scenarios, as shown in \autoref{fig:example-intersect}.
Please note that the two lanes are not connected at the intersection point, so a car cannot switch to the other lane.
\end{example}

\begin{example}[Connecting roads]
\autoref{fig:example-connect} illustrates a scenario where a car passes through a connection point and may drive into any of the connecting lanes.
\end{example}

\begin{example}[A vehicle passes two intersection points]
When a vehicle passes through two intersection points successively, there are two possible scenarios since the length of the vehicle and the distance between the intersection points are unknown, as shown in \autoref{fig:example-two-intersect}.
A long truck may cover both intersection points simultaneously, while a short car can be in between.
\end{example}

\begin{example}[Overtake on the opposite side]
TSL can use overlap to create overtaking on opposite scenarios, as shown in \autoref{fig:example-overlap}.
It contains two roads: one road has two lanes, l1 and l2, running from left to right, and the other road has one lane, l3, running from right to left, with l3 overlapping with l2.
With no lane changes allowed for c2 and c3, two scenarios are provided by TSL for c1 to overtake c2: waiting for c3 to pass before overtaking, or overtaking before c3 passes.
\end{example}


The examples above show how TSL works for scenario generation.
By combining these primitives, TSL can manage different urban road networks and create a variety of traffic scenarios.
Please refer to our source code for the detailed implementation of these examples.
\autoref{thm:scenario-existance} shows that TSL is not limited to such simple scenarios, but can theoretically handle arbitrarily complex scenarios.
More complex examples are provided in the Technical Appendix.

\subsection{Use TSL for Test Scenario Generation}

The scenarios generated can be described using a scenario description language and simulated as test cases for autonomous vehicles.
We use OpenSCENARIO DSL (OSC)\footnote{https://www.asam.net/standards/detail/openscenario-dsl/} as an example.
It is a widely accepted industry standard proposed by ASAM for describing abstract scenarios, particularly for autonomous systems testing uses.
For the syntax and semantics of OSC, please refer to its document.

A TSL scenario can be easily translated into OSC with the following steps (OSC keywords shown in \texttt{typewriter font}):
\begin{enumerate}
    \item Create a top-level \texttt{scenario};
    \item Declare all points introduced in \autoref{subsec:urban-road-network} with the \texttt{position\_3d} type;
    \item Add a \texttt{do serial} block to the top-level scenario. Then, for each scene in the TSL scenario, add a corresponding \texttt{parallel} sub-block;
    \item In each \texttt{parallel} sub-block, describe the \texttt{drive()} action of every vehicle in this scene. The OSC provides the \texttt{position} modifier for the longitudinal position (relative to another vehicle or point declared in step 2) and \texttt{lateral} for the lateral position (relative to lane boundaries).
\end{enumerate}
This translation is straightforward and can be automatized.
A complete example is provided in the Technical Appendix.
An OSC file can be further evaluated by a simulator to serve as a test case~\citep{zhang2021test,bagschik2018ontology} or used to find possible malfunctions~\citep{wang2023a2cost}.
Specific usages of OSC files are beyond the scope of this paper.

\subsection{Discussion}
TSL has numerous potential applications beyond test scenario generation.
For example, it can accomplish autonomous driving decision-making in a more interpretable manner and seamlessly integrate predictions for other vehicles into its reasoning process.
A demonstration is provided in the Technical Appendix.
It can also be used to verify the safety of control sequences for autonomous vehicles, which is a focus of our future research.

New predicates and constraints can also be introduced to extend the semantics of TSL for broader applications.
For example, predicates representing vehicle types can be added, and additional traffic rules specific to certain vehicle types can be incorporated.



\section{Conclusion}


In this paper, we propose a novel spatial-temporal logic, Traffic Scenario Logic, for modeling and reasoning of urban pedestrian-free traffic scenarios.
TSL is compatible with OpenDRIVE and has an efficient and complete spatial representation, making it well-suited for developing a human-interpretable test scenario library for autonomous vehicle safety assessment.
Our future work will involve incorporating pedestrians into scenarios and exploring additional applications of TSL.

\section*{Acknowledgments}
This work was partially supported by National Key R\&D Program of China No. 2023YFB4704500, Hunan Province Major Scientific and Technological Project No. 2024QK2001, and was partially supported by Sangfor Technologies Co., Ltd.

\bibliography{reference}

\includepdf[pages=-]{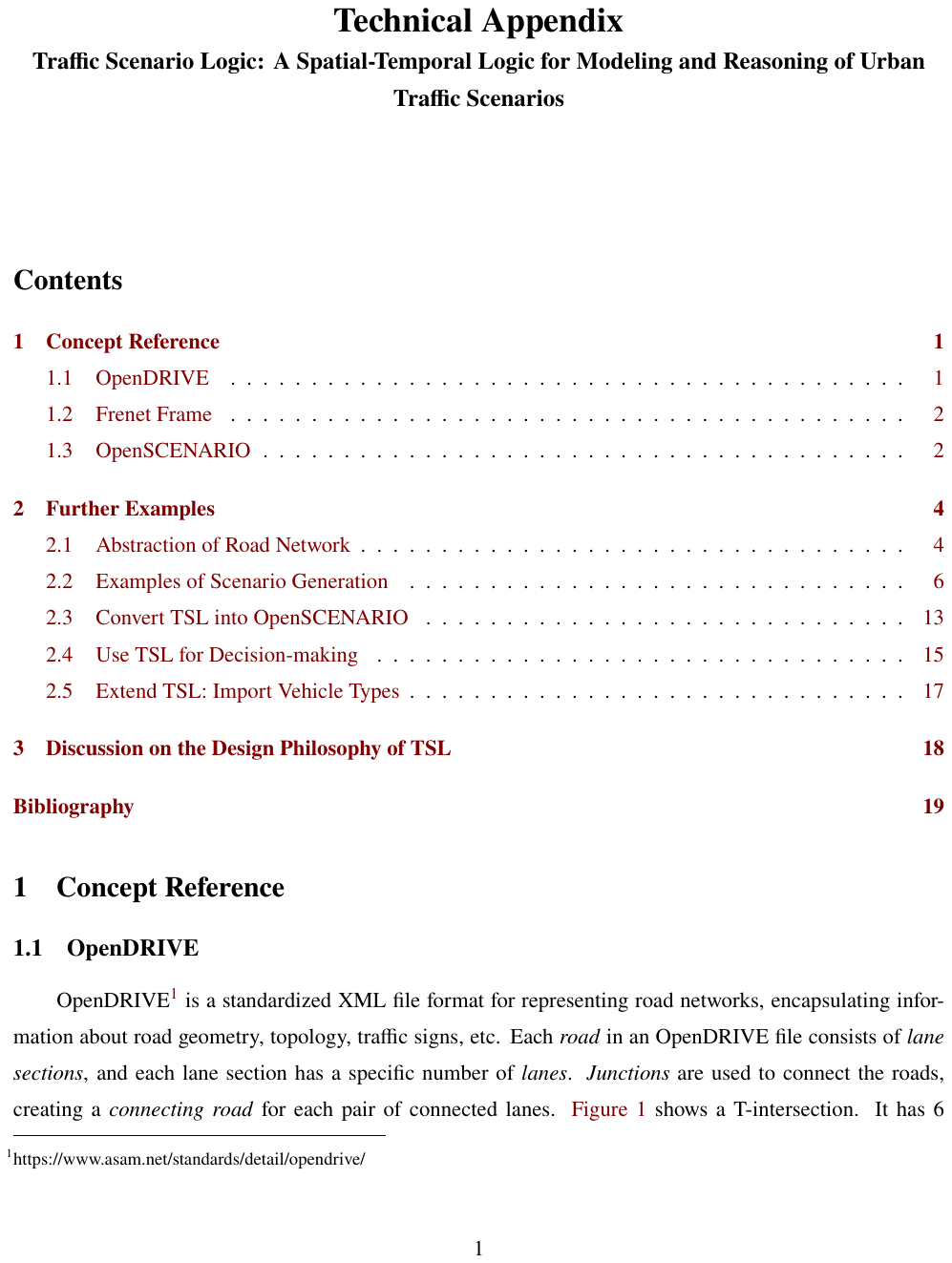}

\end{document}